# Multi-Stage Transformation and Lattice Fluctuation at AgCl-Ag Interface

*Jingshan S. Du,[†,‡,§,#] Jungwon Park,\*,[†,⊥,○] QHwan Kim,[∥] Wonho Jhe,[∥] Vinayak P. Dravid,[#] Deren Yang,\*,[‡] David A. Weitz[†,¶]*

[†] School of Engineering and Applied Sciences, Harvard University, Cambridge, Massachusetts 02138, United States

[‡] State Key Laboratory of Silicon Materials and School of Materials Science and Engineering, Zhejiang University, Hangzhou 310027, People's Republic of China

[§] Chu Kochen Honors College, Zhejiang University, Hangzhou 310058, People's Republic of China

[⊥] School of Chemical and Biological Engineering, Seoul National University, Gwanak-gu, Seoul 151-747, Republic of Korea

[○] Center for Nanoparticle Research, Institute for Basic Science (IBS), Seoul 08826, Republic of Korea

[#] Department of Materials Science and Engineering, Northwestern University, Evanston, Illinois 60208, United States

[∥] Department of Physics and Astronomy, Seoul National University, Gwanak-gu, Seoul 151-747, Republic of Korea

[¶] Department of Physics, Harvard University, Cambridge, Massachusetts 02138, United States

ABSTRACT Solid-state transformation is often accompanied by mechanical expansion/compression, due to their volume change and structural evolution at interfaces at the atomic scale. However, these two types of dynamics are usually difficult to monitor in the same time. In this work, we use *in-situ* transmission electron microscopy to directly study the reduction transformation at the AgCl-Ag interface. Three stages of lattice fluctuations were identified and correlated to the structural evolution. During the steady state, a *quasi*-layered growth mode of Ag in both vertical and lateral directions were observed due to the confinement of AgCl lattices. The development of planar defects and depletion of AgCl are respectively associated with lattice compression and relaxation. Topography and structure of decomposing AgCl was further monitored by *in-situ* scanning transmission electron microscopy. Silver species are suggested to originate from both the surface and the interior of AgCl, and be transported to the interface. Such mass transport may have enabled the steady state and lattice compression in this volume-shrinking transformation.

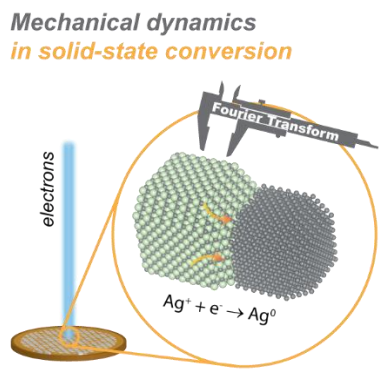

The transformation from silver halides (AgX, typical X = Cl, Br, I) to the metal Ag is a "shear transformation" that requires removal of halide atoms and reorganization of silver to form a more compact lattice. This classic solid-state transformation reaction was commonly used in the photographic reaction and was later developed as a foundation towards micro- to nano-fabrication and synthesis for various silver-based structures.[1-5] As a particular example in the emerging photocatalysis field, AgX/Ag, as a group of plasmonic semiconductor-metal hetero-nanostructures, have been synthesized from AgX solid precursors and identified to exhibit high photocatalytic efficiency under visible light.[6-10] Such a transformation reaction and the resulting interfacial structure also finds important roles in electrochemical applications. Notably, this material combination is widely used as a standard reference electrode. Therefore, the structure and dynamics at the interface between AgCl and Ag are important to understanding the charging and discharging mechanisms at the electrode, and may provide insights for other electrode materials involving analogous solid-state reactions.

To understand the transformation from silver halides to silver and the resulting interfacial structures, a method to monitor solid-state reactions with high temporal and spatial resolutions is required. Recently, *in-situ* X-ray nanodiffraction was employed to study the transformation from AgBr to Ag and revealed the existence of splitting, rotation, and lattice fluctuation of crystal grains in this reaction.[11] However, the structural origin of such dynamics needs to be further explored by direct observations in real space, since the dynamics recorded in the reciprocal space cannot be unequivocally transformed into nanoscale morphological and crystallographic information in the real space.[12]

*In-situ* transmission electron microscopy (*in-situ* TEM) has been used to provide direct observation of various chemical and physical processes involving transformation of condensed matters with multiple phases or components.[13] This technology has further assisted the design and optimization of many functional materials, such as Li-ion batteries,[14-16] semiconductor nanowires for electronics[17-19] and functional nanoparticles.[20-21] In particular, a TEM uses electrons as the illumination source, which carry high energy and can induce the conversion from AgCl to Ag. This property can enable direct observations of this reaction in high resolution using TEM.[22-24] Mechanical behaviors of materials in nanoscale has also been extensively explored thanks to the development of *in-situ* TEM nanomechanics with the aid of straining stages.[25] The understanding of plasticity and deformation mechanism in the atomic scale, for example, has been pushed forward with such technologies.[26-27] New data analysis methods such as strain mapping[28-30] have further enabled strain measurement directly from TEM images. With the emergence of these technical advances for *in-situ* TEM, the dynamic mechanical behaviors of single crystalline grains during a chemical reaction may be monitored in real time.

Here, we report the multi-stage transformation processes from AgCl to Ag and the lattice fluctuation with atomic-scale resolution at their interface, directly observed by *in-situ* TEM. The electron beam (e-beam) in TEM serves a second function in addition to imaging: high energy electrons irradiating AgCl can induce its reductive conversion to Ag and the resulting removal of Cl from the lattice.[11, 24] We monitored the growth of Ag along the AgCl-Ag atomic interface when the AgCl lattice is stable. Lattice compression and relaxation are shown to be associated with defect formation and AgCl depletion, respectively.

The AgCl-Ag heterojunction sample for *in-situ* TEM was prepared by mixing $AgNO_3$ and KCl solutions, centrifugation, and subsequent drop-casting on a TEM grid. Detailed procedures are described in the Methods Section in the Supporting Information. Light-induced reductive conversion under ambient conditions and the initial e-beam exposure during sample searching caused the formation of AgCl-Ag heterojunctions before the *in-situ* experiments. When searching for the heterojunctions, the TEM grid was screened in low magnification and low e-beam flux to minimize unwanted rapid transformation. Similar low flux was used to image an overview of typical heterojunctions before initiation of the transformation, as shown in Figure S1. Then, the flux of electrons and the magnification were raised to initiate the transformation of AgCl to Ag and were maintained constant during the observation. A suitable flux, on the order of 1 pA/cm$^2$ (normalized to a 400 kx indicated magnification, estimated from the fluorescence screen), was used for movie recording of this transformation process with sufficient signal-to-noise ratio and frame rate provided by a CCD detector. We recorded the real-time transformation processes of a heterojunction at the rate of 5 frames per second (fps) in Movie S1. TEM snapshots at different times from Movie S1 are presented in Figure 1. The heterojunction originally has a diode architecture composed of a AgCl particle, a Ag particle (Ag-P1), and their interface. As Ag$^+$ ions were continuously reduced to Ag$^0$, the interface between AgCl and Ag-P1 moved gradually into AgCl, as illustrated in stage 1. A series of dark strips were developed at the interface during stage 2 and led to the formation of a new Ag domain in stage 3. Consumption of AgCl eventually left a depleted region at the interface between two domains because of the chlorine loss from AgCl. Such a depleted region is evidenced by the low-contrast area between two materials in the TEM image at 35 s. The high spatial resolution we used in our TEM studies allows us to resolve lattice spacing of the AgCl and Ag domains. Dominant {200} lattice fringes of AgCl and {111} fringes of Ag-P1 were observed (Figure S2) and tracked. Smaller Ag particles (Ag-P2, Ag-P3) also stochastically grow, shrink, or disappear in local regions of AgCl, as shown in Figure S3 and S4.

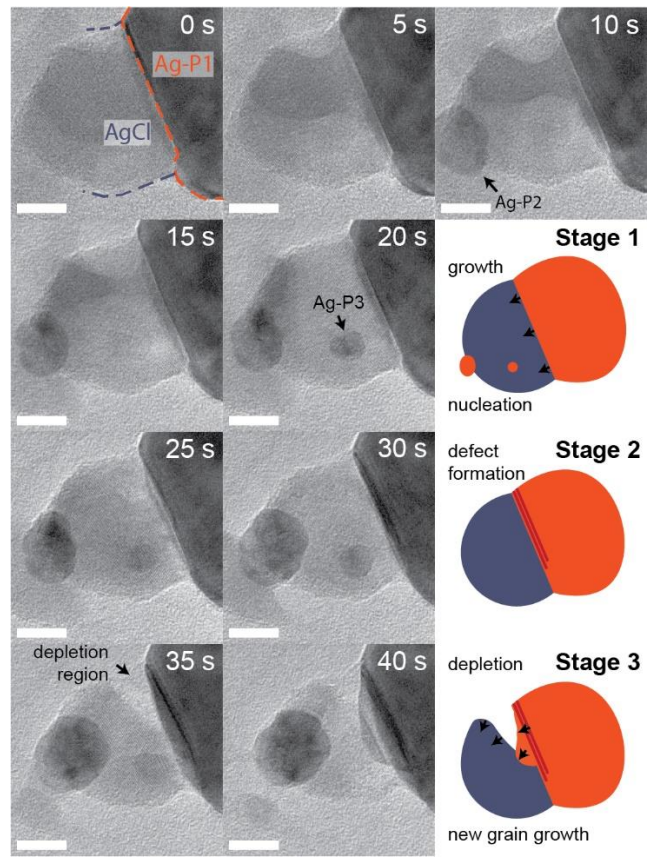

**Figure 1.** Time-series TEM images showing the transformation progress of a heterojunction and associated schematic illustrations. The initial heterojunction consists of an AgCl and an Ag (Ag-P1) part. Newly nucleated Ag Particle 2 and 3 (Ag-P2, Ag-P3) are marked at 10 s and 20 s, respectively. Scale bars are 10 nm. Ag (red) and AgCl (blue) parts are shown in the illustrations.

During the transformation process from AgCl to Ag in Figure 1, there exist fluctuations of the AgCl lattice structure. Figure 2A is a typical Fast Fourier Transform (FFT) image of the AgCl region, which clearly shows diffraction patterns along the $[0\bar{1}1]$ zone axis. After identifying instrumental noises from each TEM image, followed by measuring the centroid locations of the FFT spots (see Methods and Figure S5), average lattice spacing of both AgCl (200) and Ag (111) were calculated as shown in Figure 2B and 2C. The initial Ag (111) spacing values were used to calibrate the image scale with respect to standard lattice parameters. In this case, the average lattice spacing is mathematically defined as the weighted mean value of spatial frequencies in the reciprocal space. While the reference lattice of Ag (111) remains stable with only sub-2% fluctuations throughout the transformation, the AgCl (200) lattice spacing shows three different stages according to Figure 2C. In stage 1, the lattice spacing was generally steady around the standard value of 0.277 nm with fluctuations of *ca.* ±0.5%. It was continuously compressed to about 0.271 nm in stage 2, and then increased to pass beyond the standard value in stage 3. The orientation of the AgCl lattice also changed accordingly in these stages as shown in Figure S6.

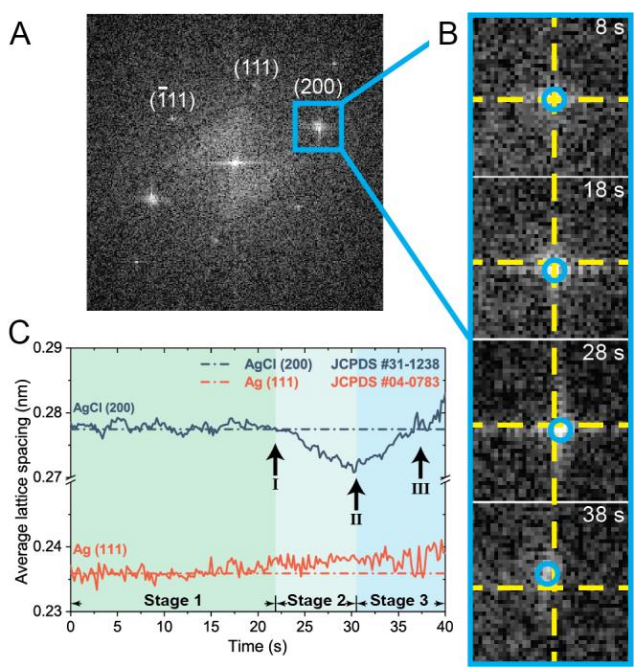

**Figure 2.** Lattice structure fluctuation during the transformation. (A) Typical FFT patterns from the AgCl region along the $[0\bar{1}1]$ zone axis. (B) Temporal fluctuation of AgCl (200) spot from the blue box in (A). Yellow dashed lines are shown for the centroid position at 8 s. Blue circles indicates the centroid position of the spots in each image. (C) Average AgCl (200) (blue) and Ag-P1 (111) (orange) lattice spacing as a function of time shows three distinct stages. Standard AgCl (200) and Ag (111) lattice spacing values are shown in dashed lines for reference.

The steady state—compression—relaxation three-stage process of the lattice fluctuation well accommodates the temporal regimes of three stages distinguished by the morphological evolution (Figure 1). To understand the underlying mechanisms, we correlated the structural changes of the AgCl-Ag interface and the lattice fluctuations in each stage.

Mechanically steady-state solid transformation occurs in stage 1. During this time, multiple monolayers of Ag were grown on the flat interface between AgCl and Ag. This interface, indicated by a black line in the color-enhanced TEM image (Figure 3A), is well aligned with the (200) lattice of AgCl. The Ag layers were continuously generated and extended along the interface, which is evidenced by the movement of their edges (black arrows) at different time. In addition, the growth pattern of silver was investigated by analyzing the contrasts in magnified TEM snapshots with corresponding schematic illustrations in Figure 3B, noting that the metallic Ag region shows significantly darker contrast than the AgCl region. With continuous reduction of $Ag^+$ from AgCl, at the initial stage (a), several segments and islands of Ag are formed at the interface of AgCl and Ag. Segmented lines in the TEM snapshot (2.0 s) indicate the forefronts of three discontinuous Ag segments, while the position of original interface is also marked for reference. As further attachment of monomers proceeds, these segments and islands connect laterally, as shown in stage (b). More islands are formed in front of the edge (2.6 s). These multiple monolayers of Ag continue to grow laterally and connect with new islands until the entire interface is filled in stage (c), also shown by the TEM image at 3.0 s. After being completely filled, the interface typically remains stable for a few seconds before new layers start to form following similar processes. It is worth pointing out that the growth does not follow exact layer-by-layer processes of monolayer attachment. Instead, the vertical height of the forming Ag layers also increases during their lateral growth. As a result, multiple monolayers are formed by both nucleation and growth in terms of atomic layers and steps. Atomic-scale nucleation takes place at multiple spots at the interface, and simultaneous growth occurs in both vertical and lateral directions.

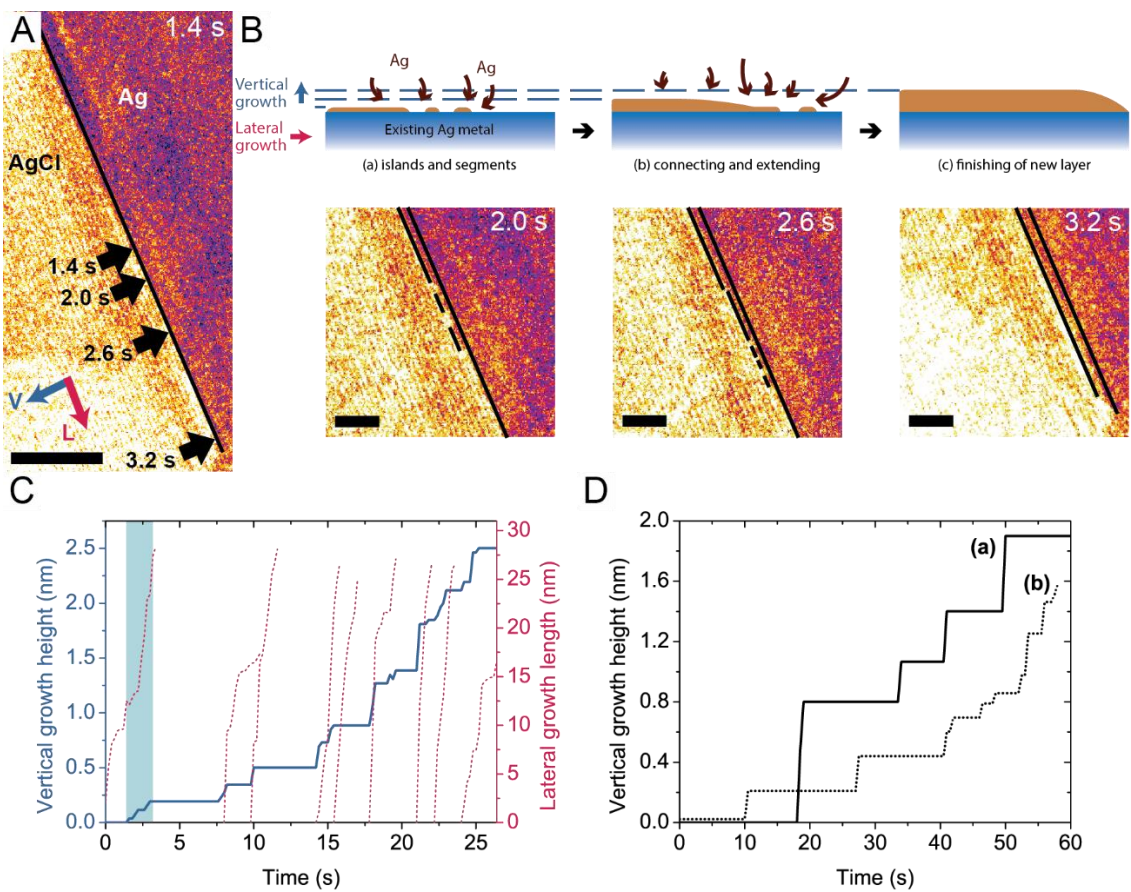

**Figure 3.** Formation of Ag in a hybrid layered manner. (A) Color-enhanced TEM image showing the AgCl-Ag interface marked by a black line. Four black arrows show the forefront position of the forming Ag layer at different time. Scale bar is 5 nm. (B) Upper row: Illustrations of the growth behavior, with both vertical (blue marks) and lateral (red marks) growth happening simultaneously. New layers (brown) form on top of the existing Ag metal (blue) with the attachment of Ag monomers from AgCl (dark red). Lower row: Magnified color-enhanced TEM images showing the forefront of the forming Ag layers [black arrows in (A)]. Black lines indicate the edges of the original and the new interface, respectively. Scale bars are 2 nm. (C) Vertical

height of the forefront (blue) and the lateral length of each layer (right, red). A blue shading indicates the time span of images in (A). (D) Vertical height of another two heterojunctions in Movie S2 (a) and Movie S3 (b) also shows stepped behaviors.

The kinetics of vertical and lateral growth is demonstrated by measuring the vertical height of the interface forefront (blue) and the lateral length of each forming layer (red) in Figure 3C. The blue curve shows stepped growth of the Ag layers in the vertical direction with respect to the interface, which resulted from the sequential formation of multiple monolayers. According to this measurement, the lateral growth rate of layers is almost two orders higher than the overall vertical growth rate. Simultaneous growth in both directions is shown by the shaded region, which corresponds to the growth events in Figure 3A. While the lateral growth progresses, the vertical height of the forefront interface rises gradually from zero to 0.19 nm in *ca.* 1.8 s, rather than jumping to the final height suddenly. Interestingly, multiple lateral growth sometimes takes place concurrently. For example, two red curves coexist from around 10 s. The growth of a second layer commenced before the completion of the first one, and eventually, these two growing layers merged into one thicker layer at 10.6 s and proceeded to fill the remaining space.

We found that the similar transformation pattern occurs at interfaces aligned with different crystallographic orientations of AgCl domains. The two curves (a) and (b) in Figure 3D are the vertical heights at the interfaces aligned with AgCl (111) and (200) in the two heterojunctions shown in Figure S7, respectively. Corresponding *in-situ* TEM movies are supplied in Movie S2 and S3. These two curves indicate that the vertical growth in different heterojunctions is stepped with uneven step sizes. Though both exhibit a discontinuity, such a growth pattern of the new phase is different from the layered or stepped growth mode recently reported in vapor phase growth of nanowires[31-32] and solid replacement of semiconductors,[17, 33] where vertical growth is accomplished by sequential growth of monolayers. In those cases, the low monomer concentration guarantees that monomer attachment preferentially takes place at atomic steps. Differently, our observations presumably demonstrate the interfacial growth in a supersaturated environment with lattice confinements. In our case, simultaneous growth of Ag to both vertical and lateral directions along the AgCl lattice is likely induced by a faster generation of $Ag^0$ than the depletion rate due to diffusion, in which supersaturated Ag monomers can initiate the formation of multiple nuclei at the interface in addition to the growth on atomic steps. Moreover, the growth of Ag is constrained by the charged framework of AgCl lattices, which leads to a layered growth behavior. This hypothesis is also evidenced by the nucleation and fast growth of Ag-P2 and Ag-P3 at around 8 s and 18 s, as shown in the TEM images in Figure 1 and the corresponding size evolution in Figure S4. Such homogeneous nucleation within the AgCl domain usually requires even higher supersaturation of monomers than heterogeneous nucleation (or growth) at an existing interface.[34-36] We emphasize that the decomposition behavior of AgCl may be dependent on the parameters of the irradiation. For instance, a higher flux and energy causes noticeable sub-second fluctuations of the contrast in Ag and AgCl domains, indicating strong diffusion inside the crystals (Figure S8).

In stage 2 and 3, the (200) lattice of AgCl underwent a dynamical fluctuation. In stage 2, the AgCl lattice was continuously compressed when the Ag domain further grew, as shown in Figure 2C. As the compression continued from *ca.* 22 s (point I) to *ca.* 30 s (point II), a series of planar defects on Ag formed near its interface with AgCl. Color-enhanced TEM images show the formation of these planar defects in Figure 4A by observing the associated dark strips. Two dark strips successively formed near the upper region of the interface at 24 s and 26 s, as indicated by blue arrows. Then, the strips extended laterally along the interface between 28 s and 30 s while the AgCl-Ag interface continued to move towards AgCl.

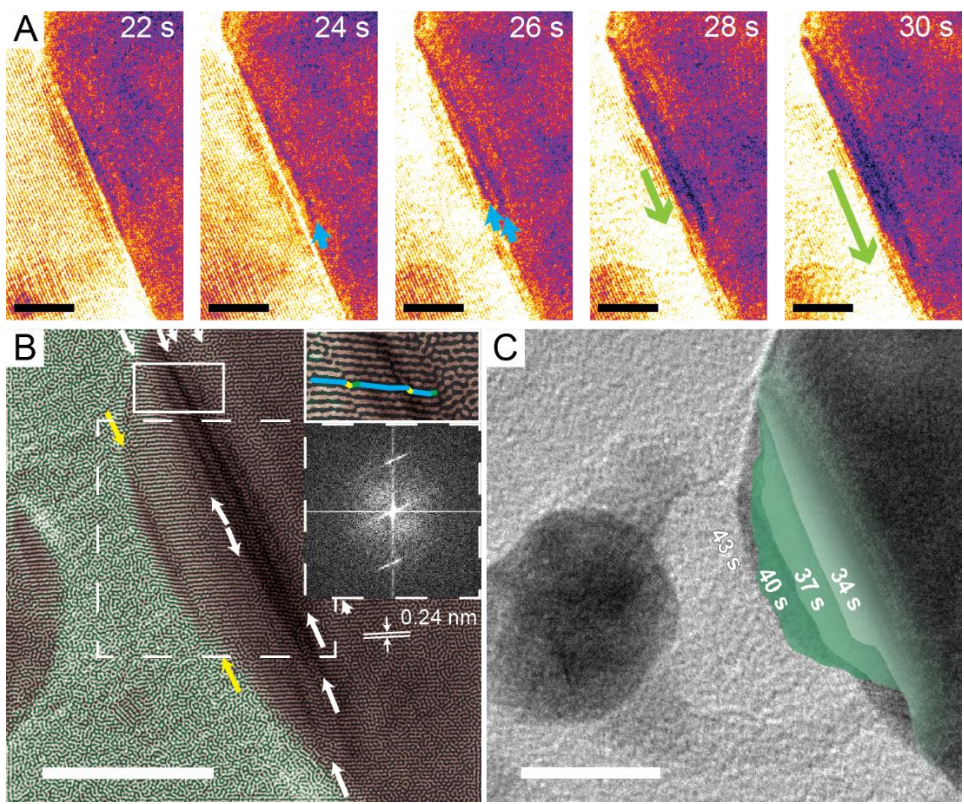

**Figure 4.** Defect formation and depletion during the transformation. (A) Time-series color-enhanced TEM images showing the formation of stacking faults. The stacking faults show darker contrast than the rest Ag domain. Scale bars are 5 nm. (B) Bandpass-enhanced high-resolution TEM image showing the stacking faults on Ag-P1. Visible interfaces due to the discontinuity of crystal structure are marked with arrows. Yellow arrows indicate twin boundaries formed in stage 3. The scale bar is 10 nm. Upper inset: magnified image of the solid box, with segmented lines showing the lattice discontinuity. Lower inset: FFT image from the original TEM image of the dashed box area. (C) TEM image shows the Ag domains grown during the depletion of AgCl. Three history lines with a time interval of 3 s are shown in green colors. The scale bar is 10 nm.

These planar defects, formed during AgCl lattice compression in stage 2, are clearly seen in the filtered high-resolution TEM image in Figure 4B at the end of the transformation. Along with the dark strips is the discontinuity of Ag lattices, indicated by white arrows. The upper inset for a magnified TEM image showing discontinuous lattice fringes, and the lower inset for a FFT image showing split patterns, both confirmed the formation of multiple stacking faults or twinning boundaries (see also Figure S9). In addition, the concurrence of defect development and AgCl lattice compression is repeatedly observed. At about 37 s, indicated as point III in Figure 2C, AgCl lattices underwent a short period of turnover in stage 3. Accordingly, an additional twin boundary was formed at this point, as shown by yellow arrows in Figure 4B.

Planar defects such as twinning and stacking faults are not rare in face-centered cubic (*fcc*) materials. This is especially true since Ag is known to have a low stacking fault formation energy among them.[37-38] Thus, planar defects in Ag crystals can be easily formed. In addition, the high monomer concentration caused by electron radiation can also facilitate the formation of stacking faults and other defects at the interface. As a result, the emergence of stacking faults inevitably alters the lattice orientation or even inserts additional partial atom planes, which exerts a pressure on the existing AgCl crystal in contact. This conclusion is also supported by the local lattice fluctuations measured in four different sub-regions of AgCl in Figure S10. In stage 2, the compression in region 2 and 4 that are adjacent to the AgCl-Ag interface, is approximately 10% larger than that measured in regions more distant from the interface (region 1 and 3). Since the AgCl domain is mechanically interacted with the carbon support underneath, the presence of this non-uniformity of lattice compression suggests that the pressure originates from the side of the interface. The clockwise rotation of AgCl (Figure S6) also indicates that a net torque exerted from the upper right side of AgCl is present. To estimate the scale of this pressure, we can assume that the AgCl nanocrystal shares the same linear elastic constants with bulk single crystals at room temperature.[39] By neglecting the strains in other directions, we calculated the stress corresponding to a normal strain of *ca.* -2.4% (with respect to the AgCl-Ag interface) to be $\sigma_{11} \approx C_{11}\epsilon_{11} \approx -1.44$ GPa. In comparison, compression of AgCl bulk crystals at $V/V_0 = 0.97$ yielded a similar pressure of 1.5 GPa.[40] These results suggest that the mechanical behavior of the single AgCl nanocrystal in our observation is in a reasonable range.

In stage 3, the compressed AgCl lattice was relaxed, as shown in Figure 2C. This change is due to the depletion of AgCl at the interface between the two materials, as shown in the TEM images in the last row of Figure 1. Consequently, the AgCl domain detached from the Ag domain, thus the (200) lattice of AgCl was gradually relieved from compression. In the same period, continuous growth of a new Ag domain was observed. Three history lines that show the surface of the growing Ag domain at different time are illustrated in Figure 4C. Contrary to the growth mode in stage 1, in which the front boundary of growing Ag formed a stepped, flat interface, a curved boundary of the Ag domain was observed in stage 3. Our previous discussions have shown that the interface morphology between AgCl and Ag in stage 1 is likely limited by the rigid and charged framework of chlorine ions,[41-42] which need to be sequentially removed during the AgCl-to-Ag transformation. In stage 3, however, Ag growth towards the depleted space is no longer subject to such a confinement. Therefore, the newly formed domain achieves the minimum surface energy by having a *quasi*-spherical morphology.

When AgCl is transformed to Ag, the volume of the whole heterojunction decreases due to the mass loss of Cl. If the reduction only involves atoms at the interface, then the receding rate of the AgCl boundary should be higher than the growth rate of Ag. Consequently, a tensile strain should be expected, similar to what is observed after about 37 s (Figure 2C), and depletion would form at the very beginning of the transformation. So how is it possible that the interface can move into AgCl steadily and a lattice compression can exist? In the previous discussions, we implicitly hypothesized that the reduced silver atoms can originate from other locations in the AgCl crystal and then diffuse to the interface. This hypothesis means that the formation rate of Ag at the interface is not necessarily lower than the receding rate of AgCl during the first two stages of our observation.

To better evaluate this hypothesized mechanism, we used *in-situ* scanning transmission electron microscopy (STEM) with simultaneous secondary electron (SE) and bright field (BF) signals to investigate the topography and structure of transforming AgCl crystals at the same time. We observed that the decomposition of AgCl can happen in two ways, as indicated by the sequential microscopy images and corresponding schematic illustrations in Figure 5. For decomposition inside the crystal, formation of a low-density region (*e.g.*, shown by the blue arrow) in the BF image without observable changes in the SE images indicates the formation of internal voids. Decomposition at the surface of AgCl is demonstrated by the recessing steps in the SE images, as the surface edges in the second image are deviated from their positions in the first image (shown by the dashed lines). Both two pathways can be potentially matched with phenomena in the *in-situ* TEM observations (Figure 1): interior decomposition may be correlated with the formation of a bright region at 10-15 s, and surface decomposition may be correlated with the diminishing of a dark region during 0-20 s. Importantly, both two observed pathways have decomposition locations away from the Ag-AgCl interfaces. These results suggest that the Ag atoms that attach to the Ag particle at the interface originate from both the interior and the surface of AgCl, and then transport to the AgCl-Ag interface, possibly due to the high Ag cation mobility in AgCl.[41-42] Note that how the chlorine ions disassemble and diffuse, and how the charge distribution on the particles evolve is not clear and requires further studies using different methods than TEM imaging. Further evidence comes from the dissolution of a small silver particle in the *in-situ* TEM experiment (Ag-P3, see Figure 1, 35-40 s; see also Figure S4 for its size evolution) when a depletion region between AgCl and Ag was formed in stage 3. This observation is in accordance with the Ostwald ripening process, in which the small particles with higher surface energy dissolve and feed monomers for the growth of larger particles. Due to the existence of such mass transport pathways of silver species in AgCl, local growth rate of Ag can be comparable or even higher than the decomposition rate of AgCl. Therefore, steady state and compression of the lattice are possible in this shear transformation, although the overall volume shrinks when changing from AgCl to Ag.

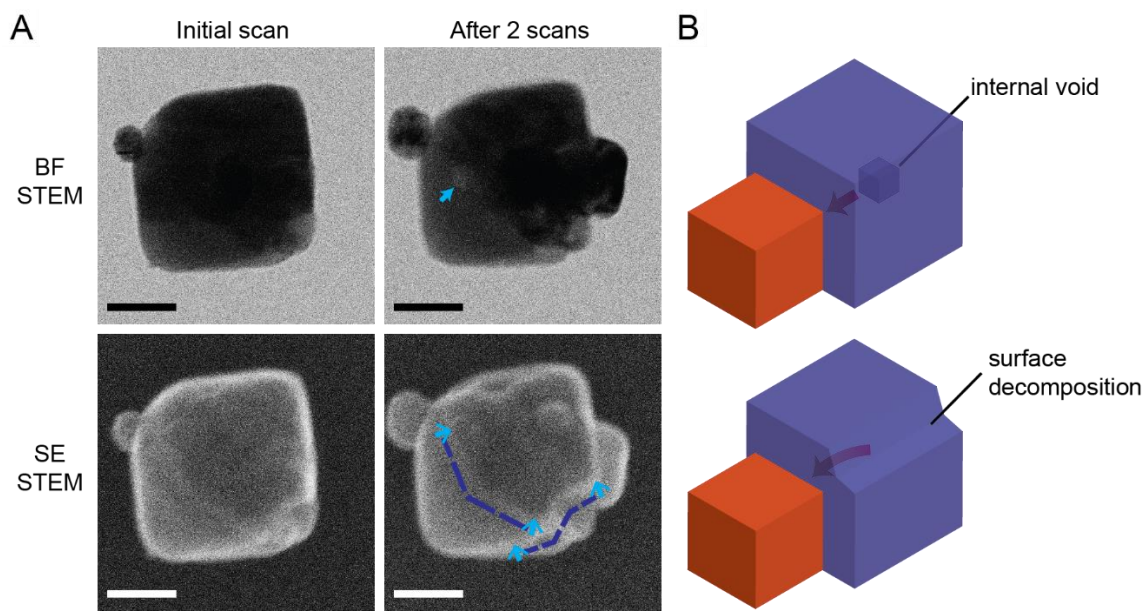

**Figure 5.** (A) Sequential STEM images showing the transformation of a AgCl particle to Ag. First row: bright field (BF) images. An arrow indicates the formation of an internal void. Second row: secondary electron (SE) images. The surface edges in the second image are deviated from the positions in the first image (shown by dashed lines). All scale bars are 50 nm. (B) Schematic illustrations for the creation of internal voids or surface decomposition, and subsequent monomer transport (arrow) to the interface of AgCl (blue) and Ag (red).

To further evaluate other possible factors that can induce lattice fluctuations, we analyzed the dynamic ranges of thermal shrinkage and grain rotation, and potential errors in the measurement. Figure S11A shows the thermal expansion and shrinkage data of *fcc* AgCl crystals, extracted from a previous X-ray diffraction (XRD) study.[43] In comparison, the dynamic range from our observations is far lower than thermal shrinkage can reach, even down to -190 °C. As for grain rotation, since a TEM image is a 2D projection, tilted crystal zone axis will lead to a multiplicity of $\cos\theta$ to the observed lattice spacing in small angles. However, this effect is very small compared to the dynamic range of our experiment. As can be seen in Figure S11B and C, the allowed crystal tilt angle is only within a few degrees to enable clear observation of lattice fringes from coherent TEM image simulation.[44] Such small tolerance angle is confirmed by analyzing the fringe-visibility band.[45] In addition, the potential measurement errors of lattice fringes are also significantly less than the dynamic range of compression–relaxation observed in this work. Detailed simulation procedures, parameters and discussions are given in the Supporting Information (Text S2–S4). Thus, the observed lattice fluctuation can be attributed predominantly to crystal compression which is associated with structural changes, rather than the two external factors.

In summary, a multi-stage transformation process from AgCl to Ag at their interfaces has been observed by *in-situ* high resolution TEM. We showed a steady state—compression—relaxation three-stage process of the lattice structure of AgCl, correlated with different structural evolution during the transformation. The presence of AgCl lattices directed formation of a flat and stepped interface as well as a *quasi*-layered growth mode in the steady state. Mass transport of Ag monomers across the AgCl crystal, either from the interior or the surface, to the AgCl-Ag interface enabled highly supersaturated growth behaviors and development of planar defects. We further showed that defect formation is associated with the compression of AgCl lattice, and the resulted strain is then relaxed due to the depletion of AgCl at the end of the transformation. Importantly, the steady state and lattice compression observed in this work will not be possible without the existence of mass transport in the converting crystal domain. We argue that in general solid-state transformations, such mass transport must be taken into consideration when judging the type of mechanical behaviors that may exist. Therefore, the results of this *in-situ* observation not only provides important details in the reduction conversion of silver halides, but can also shine light on other materials systems with similar volume change during the conversion.

## ASSOCIATED CONTENT

**Supporting Information**. Methods, additional data analysis and simulation details (Figures S1-S11), and three real-time lattice-resolution TEM movies (Movies S1-S3). This material is available free of charge via the Internet at http://pubs.acs.org.

## AUTHOR INFORMATION

**Corresponding Author**

*Email: jungwonpark@snu.ac.kr (J.P.), mseyang@zju.edu.cn (D.Y.)

**Notes**

The authors declare no competing financial interests.

## ACKNOWLEDGMENT

We would like to thank Prof. Frans Spaepen (Harvard University), Prof. Jianwei Miao (University of California, Los Angeles) and Mr. Yi Yang (National Astronomical Observatory of Japan) for helpful discussions. This work was performed in part at the Center for Nanoscale Systems (CNS) at Harvard University, supported under NSF award no. ECS-0335765; the EPIC facility of the NU*ANCE* Center at Northwestern University, supported by the SHyNE Resource (NSF NNCI-1542205), the MRSEC program (NSF DMR-1121262), the International Institute for Nanotechnology (IIN), the Keck Foundation, and the State of Illinois through the IIN. The work at Harvard was supported by the Harvard MRSEC (DMR- 1420570) and the NSF (DMR-1310266). J. P acknowledge support by IBS-R006-D1. J.S.D. acknowledges support from Zhejiang University Chu Kochen Honors College.

REFERENCES

(1) Wiley, B.; Sun, Y.; Xia, Y. Synthesis of Silver Nanostructures with Controlled Shapes and Properties. *Acc. Chem. Res.* **2007**, *40*, 1067-1076.
(2) Ballantyne, J. P.; Nixon, W. C. Selective Area Metallization by Electron-Beam Controlled Direct Metallic Deposition. *J. Vac. Sci. Technol.* **1973**, *10*, 1094-1097.
(3) Kolwicz, K. D.; Chang, M. S. Silver Halide‐Chalcogenide Glass Inorganic Resists for X‐Ray Lithography. *J. Electrochem. Soc.* **1980**, *127*, 135-138.
(4) Deng, T.; Arias, F.; Ismagilov, R. F.; Kenis, P. J. A.; Whitesides, G. M. Fabrication of Metallic Microstructures Using Exposed, Developed Silver Halide-Based Photographic Film. *Anal. Chem.* **2000**, *72*, 645-651.
(5) Schuette, W. M.; Buhro, W. E. Silver Chloride as a Heterogeneous Nucleant for the Growth of Silver Nanowires. *ACS Nano* **2013**, *7*, 3844-3853.
(6) Wang, P.; Huang, B.; Qin, X.; Zhang, X.; Dai, Y.; Wei, J.; Whangbo, M.-H. Ag@AgCl: A Highly Efficient and Stable Photocatalyst Active under Visible Light. *Angew. Chem. Int. Ed.* **2008**, *47*, 7931-7933.
(7) An, C.; Peng, S.; Sun, Y. Facile Synthesis of Sunlight-Driven AgCl:Ag Plasmonic Nanophotocatalyst. *Adv. Mater.* **2010**, *22*, 2570-2574.
(8) An, C.; Wang, R.; Wang, S.; Zhang, X., Converting AgCl Nanocubes to Sunlight-Driven Plasmonic AgCl : Ag Nanophotocatalyst with High Activity and Durability. *J. Mater. Chem.* **2011**, *21*, 11532-11536.
(9) An, C.; Wang, J.; Jiang, W.; Zhang, M.; Ming, X.; Wang, S.; Zhang, Q. Strongly Visible-light Responsive Plasmonic Shaped AgX:Ag (X = Cl, Br) Nanoparticles for Reduction of $CO_2$ to Methanol. *Nanoscale* **2012**, *4*, 5646-5650.

(10) Tang, Y.; Jiang, Z.; Xing, G.; Li, A.; Kanhere, P. D.; Zhang, Y.; Sum, T. C.; Li, S.; Chen, X.; Dong, Z.; Chen, Z. Efficient Ag@AgCl Cubic Cage Photocatalysts Profit from Ultrafast Plasmon-Induced Electron Transfer Processes. *Adv. Funct. Mater.* **2013,** *23*, 2932-2940.
(11) Huang, Z.; Bartels, M.; Xu, R.; Osterhoff, M.; Kalbfleisch, S.; Sprung, M.; Suzuki, A.; Takahashi, Y.; Blanton, T. N.; Salditt, T.; Miao, J. Grain Rotation and Lattice Deformation during Photoinduced Chemical Reactions Revealed by in situ X-ray Nanodiffraction. *Nat. Mater.* **2015,** *14*, 691-695.
(12) Sturmat, M.; Koch, R.; Rieder, K. H. Real Space Investigation of the Roughening and Deconstruction Transitions of Au(110). *Phys. Rev. Lett.* **1996,** *77*, 5071-5074.
(13) Campbell, G. H.; McKeown, J. T.; Santala, M. K. Time Resolved Electron Microscopy for in situ Experiments. *Appl. Phys. Rev.* **2014,** *1*, 041101.
(14) Huang, J. Y.; Zhong, L.; Wang, C. M.; Sullivan, J. P.; Xu, W.; Zhang, L. Q.; Mao, S. X.; Hudak, N. S.; Liu, X. H.; Subramanian, A.; *et al*. In Situ Observation of the Electrochemical Lithiation of a Single $SnO_2$ Nanowire Electrode. *Science* **2010,** *330*, 1515-1520.
(15) Liu, X. H.; Wang, J. W.; Huang, S.; Fan, F.; Huang, X.; Liu, Y.; Krylyuk, S.; Yoo, J.; Dayeh, S. A.; Davydov, A. V.; *et al.* In situ Atomic-Scale Imaging of Electrochemical Lithiation in Silicon. *Nat. Nanotechnol.* **2012,** *7*, 749-756.
(16) McDowell, M. T.; Lee, S. W.; Harris, J. T.; Korgel, B. A.; Wang, C.; Nix, W. D.; Cui, Y. In Situ TEM of Two-Phase Lithiation of Amorphous Silicon Nanospheres. *Nano Lett.* **2013,** *13*, 758-764.
(17) Chou, Y.-C.; Wu, W.-W.; Cheng, S.-L.; Yoo, B.-Y.; Myung, N.; Chen, L. J.; Tu, K. N. In-situ TEM Observation of Repeating Events of Nucleation in Epitaxial Growth of Nano CoSi2 in Nanowires of Si. *Nano Lett.* **2008,** *8*, 2194-2199.
(18) Boston, R.; Schnepp, Z.; Nemoto, Y.; Sakka, Y.; Hall, S. R. In situ TEM Observation of a Microcrucible Mechanism of Nanowire Growth. *Science* **2014,** *344*, 623-626.
(19) Chou, Y.-C.; Tang, W.; Chiou, C.-J.; Chen, K.; Minor, A. M.; Tu, K. N. Effect of Elastic Strain Fluctuation on Atomic Layer Growth of Epitaxial Silicide in Si Nanowires by Point Contact Reactions. *Nano Lett.* **2015,** *15*, 4121-4128.
(20) Niu, K.-Y.; Park, J.; Zheng, H.; Alivisatos, A. P. Revealing Bismuth Oxide Hollow Nanoparticle Formation by the Kirkendall Effect. *Nano Lett.* **2013,** *13*, 5715-5719.
(21) McDowell, M. T.; Lu, Z.; Koski, K. J.; Yu, J. H.; Zheng, G.; Cui, Y. In Situ Observation of Divergent Phase Transformations in Individual Sulfide Nanocrystals. *Nano Lett.* **2015,** *15*, 1264-1271.
(22) Hall, C. E.; Schoen, A. L. Application of the Electron Microscope to the Study of Photographic Phenomena. *J. Opt. Soc. Am.* **1941,** *31*, 281-285.
(23) Berry, C. Structure of Thin Films of Silver and Silver Bromide Substrates. *Acta Cryst.* **1949,** *2*, 393-397.
(24) Wu, Y. A.; Li, L.; Li, Z.; Kinaci, A.; Chan, M. K. Y.; Sun, Y.; Guest, J. R.; McNulty, I.; Rajh, T.; Liu, Y. Visualizing Redox Dynamics of a Single Ag/AgCl Heterogeneous Nanocatalyst at Atomic Resolution. *ACS Nano* **2016,** *10*, 3738-46.
(25) Yu, Q.; Legros, M.; Minor, A. M. In situ TEM Nanomechanics. *MRS Bull.* **2015,** *40*, 62-70.
(26) Sun, J.; He, L.; Lo, Y.-C.; Xu, T.; Bi, H.; Sun, L.; Zhang, Z.; Mao, S. X.; Li, J. Liquid-like Pseudoelasticity of Sub-10-nm Crystalline Silver Particles. *Nat. Mater.* **2014,** *13*, 1007-1012.
(27) Wang, J.; Zeng, Z.; Weinberger, C. R.; Zhang, Z.; Zhu, T.; Mao, S. X. In situ Atomic-Scale Observation of Twinning-Dominated Deformation in Nanoscale Body-Centred Cubic Tungsten. *Nat. Mater.* **2015,** *14*, 594-600.

(28) Hÿtch, M. J.; Snoeck, E.; Kilaas, R. Quantitative Measurement of Displacement and Strain Fields from HREM Micrographs. *Ultramicroscopy* **1998,** *74*, 131-146.
(29) Hÿtch, M. J.; Putaux, J.-L.; Pénisson, J.-M. Measurement of the Displacement Field of Dislocations to 0.03Å by Electron Microscopy. *Nature* **2003,** *423*, 270-273.
(30) Hüe, F.; Hÿtch, M.; Bender, H.; Houdellier, F.; Claverie, A. Direct Mapping of Strain in a Strained Silicon Transistor by High-Resolution Electron Microscopy. *Phys. Rev. Lett.* **2008,** *100*, 156602.
(31) Hofmann, S.; Sharma, R.; Wirth, C. T.; Cervantes-Sodi, F.; Ducati, C.; Kasama, T.; Dunin-Borkowski, R. E.; Drucker, J.; Bennett, P.; Robertson, J. Ledge-Flow-Controlled Catalyst Interface Dynamics during Si Nanowire Growth. *Nat. Mater.* **2008,** *7*, 372-375.
(32) Rackauskas, S.; Jiang, H.; Wagner, J. B.; Shandakov, S. D.; Hansen, T. W.; Kauppinen, E. I.; Nasibulin, A. G. In Situ Study of Noncatalytic Metal Oxide Nanowire Growth. *Nano Lett.* **2014,** *14*, 5810-5813.
(33) Wang, S.-C.; Lu, M.-Y.; Manekkathodi, A.; Liu, P.-H.; Lin, H.-C.; Li, W.-S.; Hou, T.-C.; Gwo, S.; Chen, L.-J. Complete Replacement of Metal in Metal Oxide Nanowires via Atomic Diffusion: In/ZnO Case Study. *Nano Lett.* **2014,** *14*, 3241-3246.
(34) Mer, V. K. L. Nucleation in Phase Transitions. *Ind. Eng. Chem.* **1952,** *44*, 1270-1277.
(35) Xia, Y.; Xiong, Y.; Lim, B.; Skrabalak, S. E. Shape-Controlled Synthesis of Metal Nanocrystals: Simple Chemistry Meets Complex Physics? *Angew. Chem. Int. Ed.* **2009,** *48*, 60-103.
(36) Rigorously speaking, it is more likely to be heterogeneous nucleation on carbon support versus on Ag in this specific system, whereas the energy requirement for Ag to nucleate on a different material (carbon) is always higher than on the same material (Ag).
(37) Dillamore, I. L.; Smallman, R. E. The Stacking-Fault Energy of F.C.C. Metals. *Philos. Mag.* **1965,** *12*, 191-193.
(38) Bufford, D.; Wang, H.; Zhang, X. High Strength, Epitaxial Nanotwinned Ag Films. *Acta Mater.* **2011,** *59*, 93-101.
(39) Hidshaw, W.; Lewis, J. T.; Briscoe, C. V. Elastic Constants of Silver Chloride from 4.2 to 300°K. *Phys. Rev.* **1967,** *163* (3), 876-881.
(40) Vaidya, S. N.; Kennedy, G. C. Compressibility of 27 Halides to 45 kbar. *J. Phys. Chem. Solids* **1971,** *32*, 951-964.
(41) Keen, D. A. Disordering Phenomena in Superionic Conductors. *J. Phys.: Condens. Matter* **2002,** *14*, R819.
(42) Stephen, H. Superionics: Crystal Structures and Conduction Processes. *Rep. Prog. Phys.* **2004,** *67*, 1233.
(43) Lawn, B. Thermal Expansion of Silver Halides. *Acta Cryst.* **1963,** *16*, 1163-1169.
(44) Kirkland, E. J. *Advanced Computing in Electron Microscopy*. 2nd ed.; Springer US: New York, 2010.
(45) Fraundorf, P.; Qin, W.; Moeck, P.; Mandell, E. Making Sense of Nanocrystal Lattice Fringes. *J. Appl. Phys.* **2005,** *98*, 114308.